# *In silico* modeling of the molecular structure and binding of leukotriene A4 into leukotriene A4 hydrolase

Paula B. Paz, Esteban G. Vega-Hissi, Mario R. Estrada, Juan C. Garro Martinez*

Area de Química Física, Departamento de Química, Universidad Nacional de San Luis, Chacabuco 917, San Luis, 5700, Argentina.

*Corresponding author: Tel./Fax (54) 2652 423789 int 122;
E-mail: jcgarro@unsl.edu.ar

**Abstract**

A combined molecular docking and molecular structure *in silico* analysis on the substrate and product of leukotriene A4 hydrolase (LTA4H) was performed. The molecular structures of the substrate leukotriene A4 (LTA4) and product leukotirene B4 (LTB4) were studied through Density Functional Theory (DFT) calculations at the B3LYP/6-31+G(d) level of theory in both, gas and condensed phases. The whole LTB4 molecule was divided into three fragments (hydrophobic tail, triene motif, and a polar acidic group) which were subjected to a full conformational study employing the most stable conformations of them to build conformers of the complete molecule and geometry optimize further. LTA4 conformers structures were modeled from the LTB4 minimum energy conformers. Both, protonated and deprotonated species of LTA4 and LTB4, were analyzed according to pKa values founded in the literature. Finally, a binding model of LTA4 with LTA4 hydrolase is proposed according to docking results which show intermolecular interactions that position the protonated and deprotonated ligand in the active site, in excellent agreement with experimental data.

**Introduction**

Leukotrienes (LTs) constitute a family of endogenous metabolites of arachidonic acid that are biosynthesized via the lipoxygenase pathway (1-3). These interesting compounds are a class of lipid mediators involved in the development and maintenance of inflammatory and allergic reactions (4-7). Leukotriene A4 (LTA4), an unstable alkyl epoxide formed from the immediate precursor 5-HPETE via 5-lipoxygenase (5-LO) (8), is converted to leukotriene B4 (5$S$,12$R$-dihydroxy-6$Z$,8$E$,10$E$,14$Z$-eicosatetraenoicacid; LTB4, Figure 1), by stereoselective hydratation of LTA4 hydrolase (LTA4H) (9). LTB4 is a potent pro-inflammatory mediator implicated in the pathogenesis of a number of diseases including inflammatory bowel disease (IBD), psoriasis, rheumatoid arthritis and asthma, and plays an important role in immunological responses, chronic obstructive pulmonary disease (COPD) and atherosclerosis (4,6,10-12).

On the other hand, LTA4H is a bifunctional zinc metalloenzyme which catalyzes the rate limiting step in the production of LTB4 (13). The X-ray crystal structures of LTA4H in complex with different inhibitors have been obtained by several authors (17-19). According to several investigations, the metal site is located next to the putative active site bound to His295, His299, and Glu318 (14-18).

In the present paper, we show an analysis of the different interactions of LTA4 into LTA4H through molecular docking study. Molecular and electronic structure information of LTA4 and LTB4, obtained from an exhaustive conformational analysis, was used to set an initial conformation for the docking study.

There are few theoretical studies about the molecular structure of LTB4. One of them was performed by Brasseur at el. using semiempirical methods (19). They present a computational description of the conformation of a pair of two isomeric molecules (6-cis and 6-trans-leukotriene B4) forming a complex with one calcium ion. Recently, Catoire at el. have obtained a highly constrained seahorse conformation when it is bound to leukotriene receptor BLT2 by $^1$H-RMN (20). However, there has been no *ab initio* and/or DFT studies that analyzes the conformation and electronic structure of LTA4 and LTB4.

Therefore, we believe that results of this work are of great contribution especially to elucidate: a) the structural similitude between LTB4 and its precursor LTA4, b) the conformational changes of LTA4 in gas phase, condensed phase and when it is bounded to the enzyme, c) the active site of LTA4H where LTA4 binds and compare it with the interaction site of inhibitors (11-13) and thus, to obtain important information for future development of new inhibitors of LTB4 formation.

**Material and Methods**

**Conformational study**

Relaxed potential energy curves (PECs) were performed at HF/3-21G, HF/6-31+G(d) and B3LYP/6-31+G(d) levels of theory (21-23), while potential energy exploratory surfaces (PESs) were carried out at HF/3-21G level of theory. The conformers were then optimized at a higher level and frequency calculations were carried out to confirm minimum energy conformers checking the absence of imaginary frequencies. Solvent effect was evaluated through geometry optimizations using the integral equation formalism polarizable continuum model (IEF-PCM) (24,25) at B3LYP/6-31+G(d) level of theory. All calculations were performed using the Gaussian 03 computational program (26).

**Molecular Docking**

Docking calculations were performed using Autodock Vina software (27) in order to propose a model of the LTA4-LTA4H complex. The crystal structure of LTA4H used in our study was obtained from Protein Data Bank (28) (PDB accession code: 3CHO (19)). The original ligand, 2-amino-N-[4-(phenylmethoxy)phenyl]-acetamide and water molecules in the crystal structure were removed. LTA4 initial conformation was obtained from the conformational study. A torsional restriction was applied for single bonds of the conjugated system of the triene motif fragment setting them as fixed, i.e., not able to rotate. The box was defined to include completely the proposed binding site and standard parameters for the docking calculation were used except for exhaustiveness for which a value of 100 was used.

The lowest binding energy docking complex was analyzed with NCIPlot program to identify and map noncovalent interactions using the promolecular densities and its derivatives. These interactions are nonlocal and manifest in real space as low-gradient isosurfaces with low densities which are interpreted and colored according to the corresponding values of sign($\lambda_2$)$\rho$ (29).

**Results and Discussions**

**Conformational study for LTB4**

To obtain the structure of LTA4 for the docking study, we previously performed a conformational analysis on LTB4, which optimized structure was used to propose the initial conformation of LTA4 for the molecular docking study.

Several papers agree that LTB4 is a cis-trans (or E-Z) isomers (5*S*,12*R*-dihydroxy-6*Z*,8*E*,10*E*,14*Z*-eicosatetraenoicacid) (19,20,30) and that LTB4 show conformational restrictions in the double bonds configuration of LTB4 (6Z,8E,10E,14Z). Despite the restrictions of double bonds, the molecule of LTB4 has 13 rotational bonds, and if these dihedrals were affected by systematic 120º changes, more than 1.500x10$^3$ conformations could be obtained. To avoid this large number of possible conformations, a systematic analysis was realized in a stepwise manner on three different important fragments of the LTB4 easily recognizable: a) hydrophobic tail b) a triene motif and c) a polar group (Figure 2). Our systematic study includes a severe analysis of each fragment which was selected based on its chemical characteristics. Molecular fragmentation have been used in previous work of our group given excellent results in the study of large molecules (31,32).

*Hydrophobic tail*

This fragment is composed by a chain of methylens units which is bound to the triene fragment through a double bond. Conformational study of this fragment consisted in the analysis of the four dihedral angles presented in Figure 2a. Rotating dihedrals $\phi_1$ and $\phi_4$ two potential energy curves were obtained (Figure 3a and b). The plots present the global energy minimum around 180º for $\phi_1$ and two local minima next to 120º and -120º with the same energy for $\phi_4$. PES of the type E=f($\phi_2$,$\phi_3$) shows a global minimum with dihedral values of about $\phi_2$=180º and $\phi_3$=180º. This is show in supplementary material, Figure 1S.

We have found that the hydrophobic tail fragment has high symmetry and an elongated minimum energy conformation with dihedral values next to 180º for $\phi_1$, $\phi_2$, $\phi_3$, and 120º or -120º for $\phi_4$. However, the low rotational barriers (less than 3 kcal.mol$^{-1}$) indicate that this fragment has a high conformational flexibility.

*Triene motif*

Although the conjugated triene system confers the fragment a high rigidity, there are dihedral angles that deserve to be analyzed: $\phi_5$, $\phi_6$ and $\phi_7$, Figure 2. In contrast to these dihedrals angles, the values of dihedral $\phi_8$ depend largely on the fragment with the acidic function. So, we believe that the analysis of this dihedral has no practical meaning at this point.

Dihedral angle $\phi_5$ was studied through a PEC, Figure 3c. The global energy minimum has the double bond in opposite position respect to the OH group, next to 120º, where no interaction occurs, but probably torsional tension is relieved. PES of the type E=f($\phi_6$, $\phi_7$) shown in supporting information as Figure 1S, presents the global minimum on the region of $\phi_6 \approx 180º$ and $\phi_7 \approx -120º$. However, there are local minima with rotational barriers of about 3 kcal.mol$^{-1}$ and energy values close to global energy minimum. The low rotational barrier confers a relatively flexibility, that would only be restrained in the complete LTB4 molecule by steric repulsion.

*Polar (Acidic) group*

This fragment is characterized by an OH group over a quiral carbon and a carboxylic group, both separated by three methylene groups. Due to the methylene chain, the fragment presents a high flexibility in the dihedral angles $\phi_9$, $\phi_{10}$, $\phi_{11}$ and $\phi_{12}$. A similar analysis to the others fragment was performed to the dihedral angles of the present fragment, briefly: PECs were made for dihedrals $\phi_9$ and $\phi_{12}$, while dihedrals $\phi_{10}$ and $\phi_{11}$ were analyzed through PES.

In Figure 3d is identified the zone of the highest energy ($\phi_9 \approx 0º$, repulsion zone) and zones of minimum energy ($\phi_9 \approx 180º$, global minimum energy and $\phi_9 \approx 60º$, $\Delta E = 0.1$ kcal.mol$^{-1}$). PES obtained from $\phi_{10}$ and $\phi_{11}$, plotted in Figure 1S, shows that the global energy minimum is localized next to 60º for both, $\phi_{10}$ and $\phi_{11}$. The low rotational barriers between any of the two minima indicate the high flexibility of the fragment.

Furthermore, rotation of dihedrals $\phi_{12}$ and $\phi_{13}$ produces a series of conformers with intramolecular hydrogen bond interactions. Consequently, a NBO analysis was performed on these conformations. In addition, taking into consideration the values of pKa for leukotrienes range from 3 to 5 (33,34), deprotonated species should also be studied. Table 1 presents the three potential conformations for acidic group in both species obtained from DFT calculations. The total energy indicates that both species of the first conformer are the most stable conformations. The interaction energy E(2) between donor and acceptor orbitals show an effective interaction between the oxygen lone pairs (LP) and the sigma antibonding orbitals ($\sigma$*) of O–H bond which is higher in the deprotonated species (no charge transfer interaction was founded in the last conformer for the anionic species). Although the third acidic conformer presents a higher charge transfer stabilization energy, steric restraints could disfavor this conformation.

*Molecular structure for LTB4*

Using the minimum energy conformers of the three fragments we proposed an initial molecular structure for LTB4. However, its stable conformation could be the same or different to structure proposed. Figure 4 shows the superposition of the most stable conformer and molecular structure obtained from the fragmentary study. The principal difference of both structures is the values of dihedral angle $\phi_8$. The value proposed for $\phi_8$ in the initial structure was of 180º (Figure 4, in dark gray) while full optimized value is -125º (Figure 4, in light gray) giving rise to a $\Delta$E between both structures of about 35 kcal.mol$^{-1}$. This result is expected because dihedral $\phi_8$ was not analyzed in the fragmentary analysis. Consequently, dihedral value was modified when the whole molecule was fully optimized.

**Molecular structure for LTA4**

LTB4 shows a high structural similarity to its precursor the leukotriene A4. The most important structural difference resides in the presence of an epoxy group in LTA4 molecule. This epoxy group is hydrolyzed and converted into an OH group in the enzymatic step that involves the formation of LTB4 (13-18).

To obtain a stable conformation for LTA4, which can be used in molecular docking study, we suggested an initial conformation using the dihedral values of LTB4. Hydrophobic tail dihedral initial values were taken from the most stable conformation of LTB4. In addition, literature suggests a 7E, 9E, 11Z, 14Z configuration for the unsaturated part of LTA4 (8). However, dihedrals $\theta_1$ and $\theta_2$ of LTA4, Figure 5, are involved in the orientation of the epoxy group which plays a key role in enzyme binding. Consequently, to perform a complete analysis, these important dihedrals were studied.

A full optimization carried out for the four possible conformers gave low energy differences between all conformers, Figure 2S in supporting information. Solvent effect analysis indicates that the stable conformations obtained *in vacuo* are not altered by the presence of the solvent.

**Molecular Docking**

A molecular docking study was performed to propose a binding model and interpret the different interactions between the ligand and the enzyme. Although the reaction mechanism proposed by Thunnissen et al. (14) involves the deprotonated species of LTA4, both, the acidic and anionic species of LTA4, were subjected to docking studies.

Nine possible binding complexes were obtained for the anionic species of LTA4. The most stable one, i.e. the one with lowest $\Delta G$ of binding (about -9.5 kcal.mol$^{-1}$), is shown in Figure 6. The model reveals that the carboxylate group interacts with Arg563 and is close to Lys565; the epoxide is positioned with the oxygen atom coordinating with $Zn^{2+}$ cation and forming a square based pyramidal metal complex with His 295, His 299 and Glu 318 (14); the hydrophobic tail is buried in a hydrophobic pocket; and Asp375 is oriented toward the triene group where it would control the position of a water molecule and therefore the stereospecific insertion of the 12*R*-hydroxyl group of LTB4 as is proposed in the literature according to the reaction mechanism presented (14-18).

The most stable docking complex of the enzyme and the acidic species of LTA4 presents the substrate located in the same cavity and with almost the same orientation that was founded for the anionic one (Figure 6). Moreover, ΔG of binding is only 0.1 kcal.mol$^{-1}$ higher.

Figure 6 also shows a complex web of weak noncovalent interactions between the substrate and the active site of the enzyme (including the Zn$^{2+}$ cation) where van der Waals and hydrophobic interactions dominate by far (green surfaces). Moreover, interaction analysis reveals a hydrogen bond between the carboxylate group of LTA4 and Arg563. In the proposed binding model LTA4 do not interact with Lys565. However, we expect that the dynamic behavior of both, the substrate and the enzyme, leads to the formation of a hydrogen bond (or more probably a salt bridge) between LTA4 and the positively charged residue Lys565 due to it is close to Arg563 in the binding site.

LTA4 minimum energy conformation from quantum analysis shows differences with enzyme-bounded conformation (Figure 7). However, it is well known that flexible molecules change their conformation upon binding to a protein (35), thus, it is not surprising our finding if we take into account the high degree of flexibility of LTA4 molecule.

**Conclusions**

We obtained from conformational analysis the stable structures for the substrate and product of the enzyme LTA4H using Hartree Fock and DFT calculations. Most stable conformation of LTB4 shows an elongated structure with an intramolecular interaction in the polar fragment analyzed through NBO analysis. Based in the structural data of LTB4 four possible conformations of LTA4 were proposed. We founded that conformer IV is the most stable in gas and solvated phases with a ΔE of 1 kcal.mol$^{-1}$ less than the other conformers.

Molecular docking results provided an interaction model of LTA4 with LTA4H in excellent correlation with the experimental data: the substrate fits in the active site of the enzyme arranging its functional groups according to the reaction mechanism proposed by several authors (14-17). Both species of LTA4, protonated and deprotonated, bind to the same site with proper orientation of the key groups for the enzymatic reaction. The substrate is held in the active site by multiple weak attractive interactions along its surface and a hydrogen bond to Arg563.

Thus, the *in silico* model of LTA4-LTA4H complex obtained from molecular docking in combination with a conformation analysis of substrate and product provides insights into the development of new and most efficient inhibitors of LTB4 formation.

**Acknowledgements**

This work was supported by Consejo Nacional de Investigaciones Científicas y Técnicas (CONICET) project PIP11220100100151 and Universidad Nacional de San Luis (UNSL).

**Figure Legends**

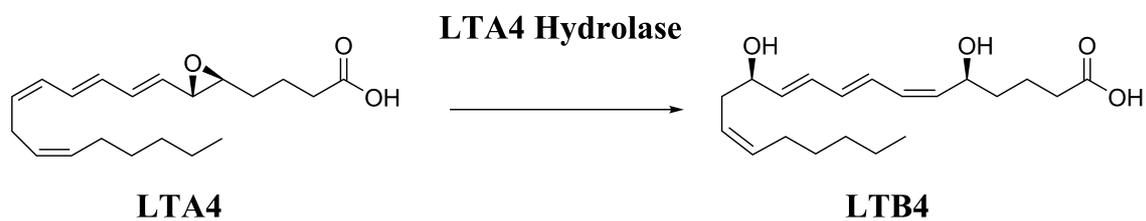

**Figure 1:** Reaction catalyzed by LTA4H. Chemical structures of LTA4 and LTB4.

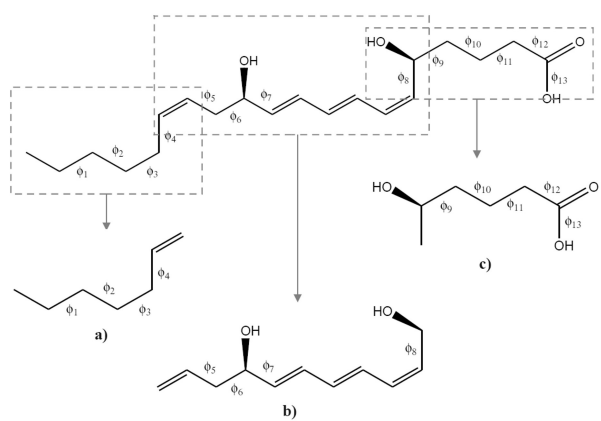

**Figure 2:** Structure of LTB4 (5*S*,12*R*-dihydroxy-6*Z*,8*E*,10*E*,14*Z*-eicosatetraenoicacid). Scheme of fragments of LTB4 and dihedral studies: a) hydrophobic tail, b) triene motif, c) polar group.

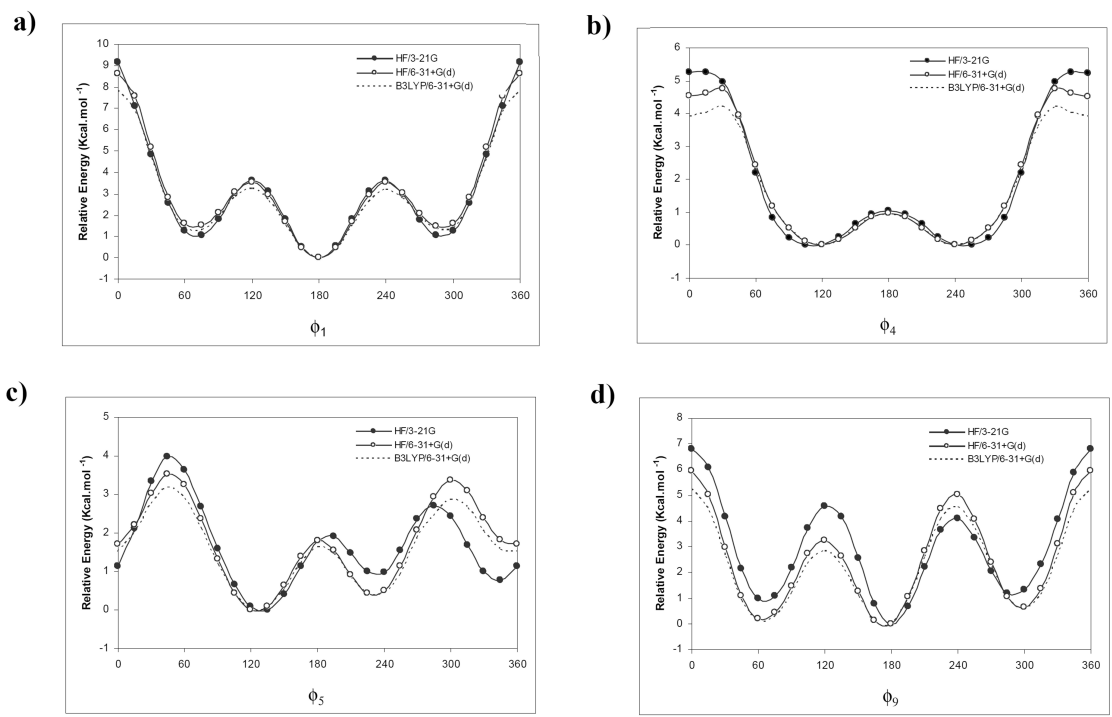

**Figure 3:** Curves energy potential for; a) and b) hydrophobic tail, c) triene motif and d) acidic part.

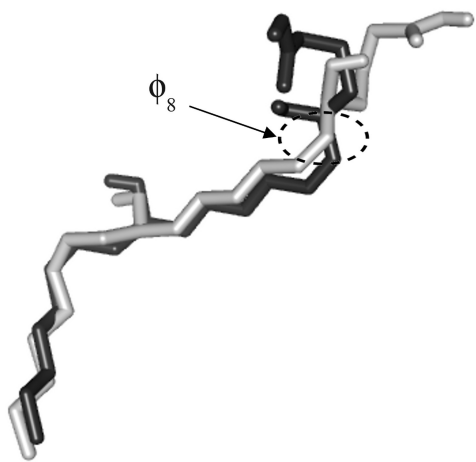

**Figure 4:** Overlap of the most stable conformer of LTB4 (light gray sticks) and molecular structure obtained from the fragmentary study (dark gray sticks).

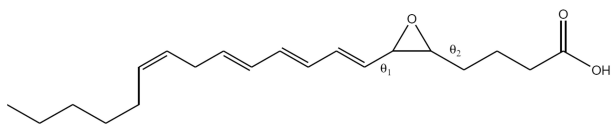

**Figure 5:** Molecular structure for LTA4 (5S,6S-epoxy-7E,9E,11Z,14Z-eicosatetraenoic acid).

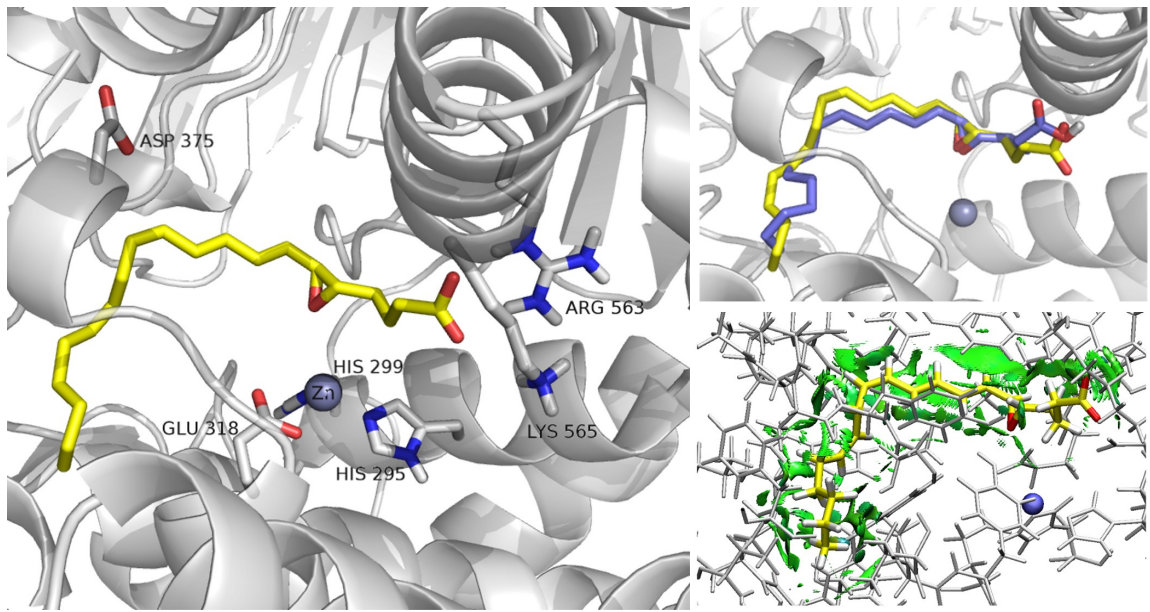

**Figure 6:** Model of LTA4 anion bounded to LTA4H. Docked complex showing the side chains of the amino acids involved in the reaction mechanism proposed (left panel). Combined docking results of acidic and anionic species of LTA4 (upper right panel). Noncovalent interactions of the LTA4 anion within the binding pocket of LTA4H (lower right panel). Blue indicates strong attractive interactions and green van der Waals (weak attractive) interactions.

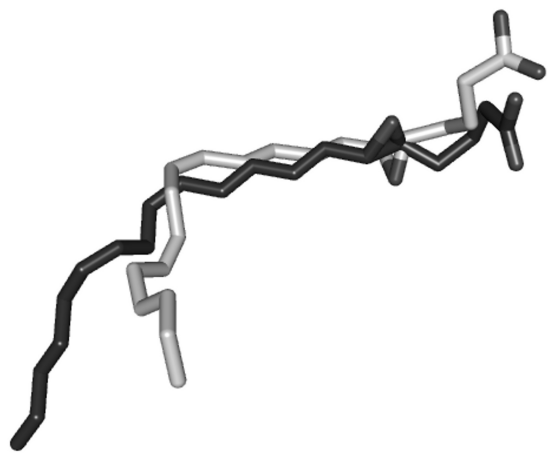

**Figure 7:** Overlap of the most stable conformer of LTA4 (dark gray sticks) and molecular structure obtained from the molecular docking study (light gray sticks).

**Table 1:** Conformational and electronic dates obtained by rotation of dihedrals $\phi_{12}$ and $\phi_{13}$ for polar fragment.

| | Form | $\phi_{12}$ | $\phi_{13}$ | $E(2)^a$ kcal.mol$^{-1}$ | Total Energy$^b$ (Hartree) | $\Delta E$ kcal.mol$^{-1}$ |
|---|---|---|---|---|---|---|
| 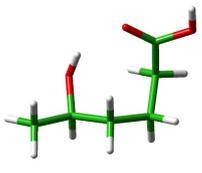 | Protonated | -77.70° | -178.70° | LP(1) 2.06 LP(2) 5.34 | -461.5722022 | 0.0 |
| | Deprotonated | -77.84° | - | LP(1) 6.98 LP(2) 34.67 | -461.0349725 | |
| 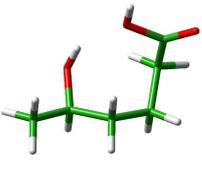 | Protonated | 86.96° | 179.52° | LP(1) 2.92 | -461.5698006 | 1.50 |
| | Deprotonated | 100.69° | - | LP(1) 6.98 LP(2) 34.67 | -461.0349725 | |
| 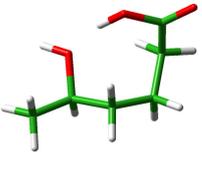 | Protonated | 94.72° | 3.01° | LP(2) 16.83 | -461.5702739 | 1.21 |
| | Deprotonated | 15.06 | - | Not founded | -461.0181532 | |

$^a$ $E(2)$ means energy of hyper conjugative interaction (stabilization energy). LP: Lone Pair

$^b$ Total energy of conformer in Hartree (1 hartree = 627.5095 kcal.mol$^{-1}$).

**Table 2:** Energy and structural dates for the conformers of LTA4.

| | HF/6-31+G(d) | | | B3LYP/6-31+G(d) | | | IEF-PCM/ B3LYP/6-31+G(d) | | |
|---|---|---|---|---|---|---|---|---|---|
| | $\theta_1$ | $\theta_2$ | $\Delta E$ (kcal.mol$^{-1}$) | $\theta_1$ | $\theta_2$ | $\Delta E$ (kcal.mol$^{-1}$) | $\theta_1$ | $\theta_2$ | $\Delta E$ (kcal.mol$^{-1}$) |
| I | -72.58 | 93.46 | 2.25 | -66.52 | 92.92 | 2.33 | -63.82 | 93.57 | 2.19 |
| II | -73.72 | -148.45 | 0.62 | -67.36 | -147.62 | 1.00 | -64.25 | -152.00 | 1.65 |
| III | 148.23 | 92.08 | 1.69 | 151.11 | 91.49 | 1.40 | 153.48 | 91.35 | 0.60 |
| IV | 148.03 | -149.07 | 0 | 150.28 | -149.90 | 0 | 154.00 | -153.16 | 0 |

**Supporting Information**

Additional Supporting Information may be found in the online version of this article:

**Figure 1S:** Potential energy surfaces: A) E=f($\phi_3,\phi_4$) hydrophobic tail, B) E=f($\phi_6,\phi_7$) triene motif, C) E=f($\phi_{10},\phi_{11}$) polar group.

**Figure 2S:** Stable conformers for LTA4 at B3LYP/6-31+G(d) levels of theory.

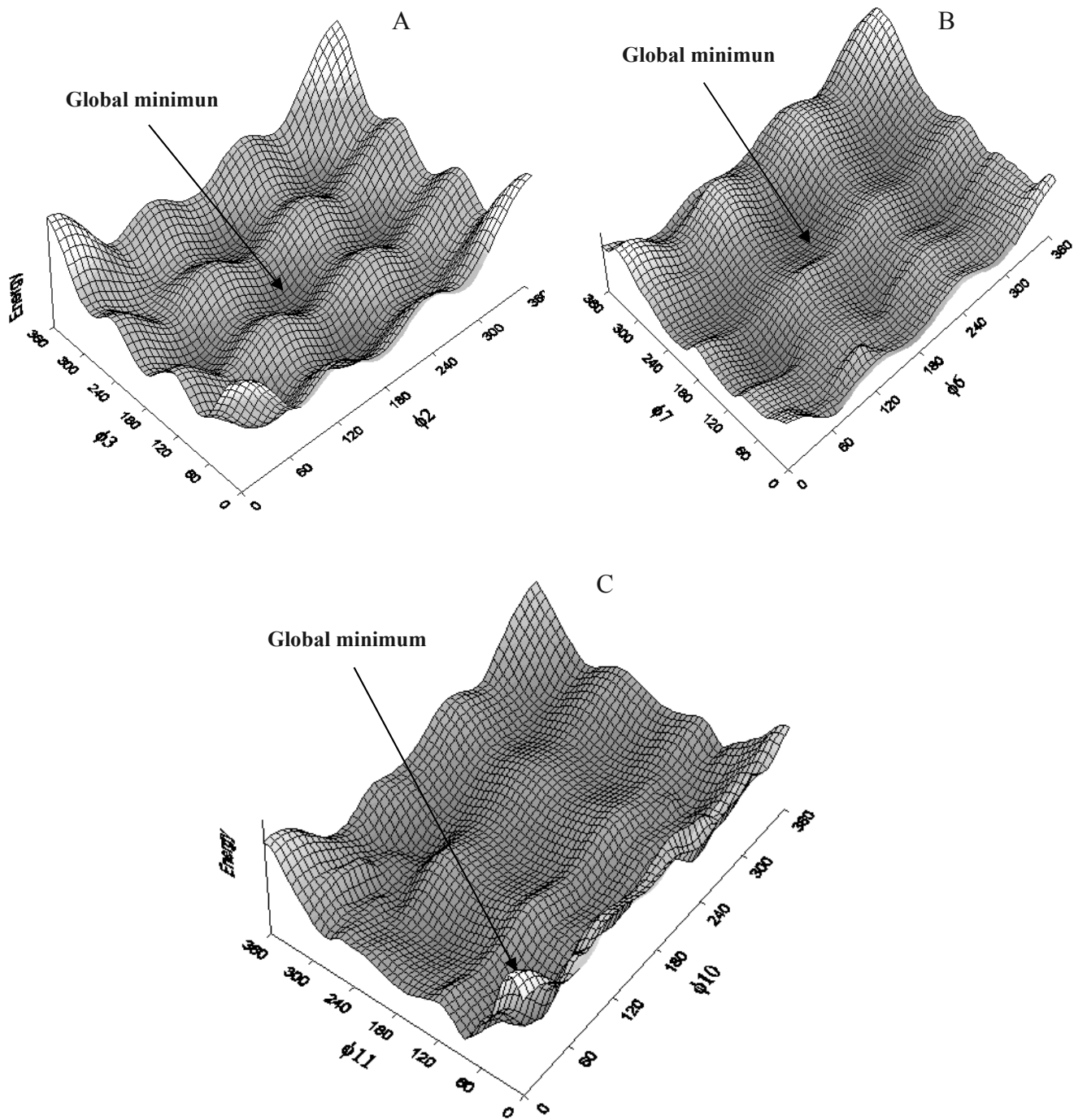

**Figure 1S:** Potential energy surfaces: A) E=f($\phi_3,\phi_4$) hydrophobic tail, B) E=f($\phi_6,\phi_7$) triene motif, C) E=f($\phi_{10},\phi_{11}$) polar group.

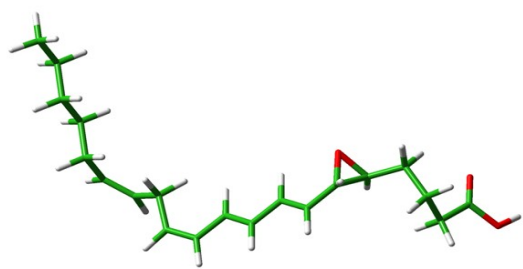

Conformer I

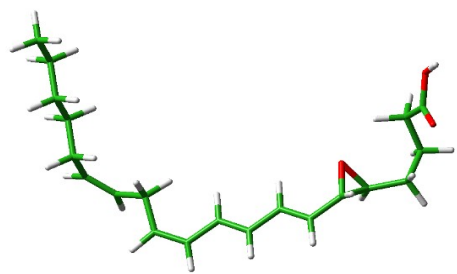

Conformer II

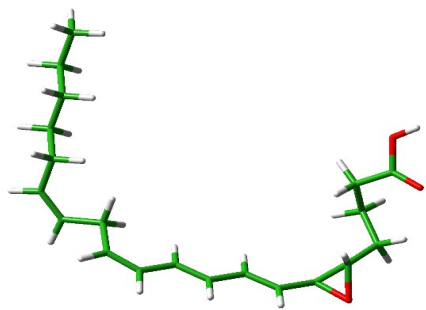

Conformer III

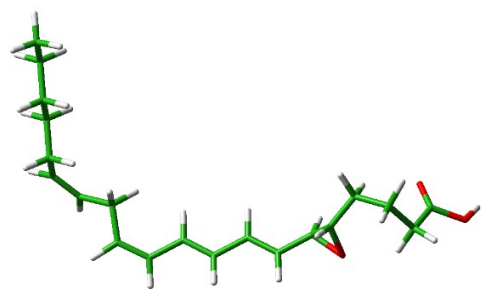

Conformer IV

**Figure 2S:** Stable conformers for LTA4 at B3LYP/6-31+G(d) levels of theory.